\documentclass[twocolumn]{emulateapj}
\usepackage{apjfonts}

\slugcomment{}

\begin{document}

%% LaTeX will automatically break titles if they run longer than
%% one line. However, you may use \\ to force a line break if
%% you desire.

\title{Deep multi-telescope photometry of NGC 5466. I. Blue Stragglers and binary systems.}
%% Use \author, \affil, and the \and command to format
%% author and affiliation information.
%% Note that \email has replaced the old \authoremail command
%% from AASTeX v4.0. You can use \email to mark an email address
%% anywhere in the paper, not just in the front matter.
%% As in the title, use \\ to force line breaks.

\author{G. Beccari\altaffilmark{1},
E. Dalessandro\altaffilmark{2},
B. Lanzoni\altaffilmark{2},
F. R. Ferraro\altaffilmark{2},
A. Sollima\altaffilmark{3},
M. Bellazzini\altaffilmark{3},
P. Miocchi\altaffilmark{2}}

%% Notice that each of these authors has alternate affiliations, which
%% are identified by the \altaffilmark after each name.  Specify alternate
%% affiliation information with \altaffiltext, with one command per each
%% affiliation.

\affil{\altaffilmark{1} European Southern Observatory, Karl--Schwarzschild-Strasse 2,
85748  Garching bei M\"unchen, Germany, gbeccari@eso.org}
\affil{\altaffilmark{2} Dipartimento di Fisica e Astronomia, Universit\`a 
degli Studi di Bologna, via Ranzani 1, I--40127 Bologna, Italy}
\affil{\altaffilmark{3} INAF--Osservatorio Astronomico 
di Bologna, via Ranzani 1, I--40127 Bologna, Italy}
%% Mark off your abstract in the ``abstract'' environment. In the manuscript
%% style, abstract will output a Received/Accepted line after the
%% title and affiliation information. No date will appear since the author
%% does not have this information. The dates will be filled in by the
%% editorial office after submission.

\begin{abstract}
We present a detailed investigation of the radial distribution of blue
straggler star and binary populations in the Galactic globular cluster
NGC 5466, over the entire extension of the system. We used a
combination of data acquired with the ACS on board the Hubble Space
Telescope, the LBC-blue mounted on the Large Binocular Telescope, and
MEGACAM on the Canadian-France-Hawaii Telescope. Blue straggler stars
show a bimodal distribution with a mild central peak and a quite
internal minimum. This feature is interpreted in terms of 
 a relatively young dynamical age
 in the framework of the ``dynamical clock'' concept proposed by Ferraro
  et al. (2012). The estimated fraction of binaries is $\sim 6-7\%$ in
  the central region ($r<90\arcsec$) and slightly lower ($\sim 5.5\%$)
  in the outskirts, at $r>200\arcsec$. Quite interestingly, the
  comparison with the results of \citet{mil12} suggests that also
  binary systems may display a bimodal radial distribution, with the
  position of the minimum consistent with that of blue straggler
  stars.  If confirmed, this feature would give additional support to
  the scenario where the radial distribution of objects more massive
  than the average cluster stars is primarily shaped by the effect of
  dynamical friction.  Moreover, this would also be consistent with
the idea that the unperturbed evolution of primordial binaries could
be the dominant BSS formation process in low-density environments.
\end{abstract}


\section{Introduction} 
\label{intro}
Galactic globular clusters (GCs) are dynamically active systems, where
stellar interactions and collisions, especially those involving
binaries, are quite frequent \citep[e.g.][]{hut92,me97} and 
generate populations of exotic objects, like X-ray binaries, millisecond pulsars,
and Blue straggler stars (BSSs; e.g.,
\citealp{paresce92,ba95,bellazz95,fe95},
2009;\citealp{ransom05,pooley06}). Among these objects, BSSs are the
most abundant and therefore may act as a
crucial probe of the interplay between stellar evolution and stellar
dynamics \citep[e.g.][]{ba95}.  They are brighter and bluer (hotter)
than main sequence (MS) turnoff (TO) stars and are located along an
extrapolation of the MS in the optical color-magnitude diagram
(CMD). Thus, they mimic a rejuvenated stellar population, with masses
larger than those of normal cluster stars ($M\gtrsim 1-1.2 M_\odot$;
this is also confirmed by direct measurements; \citealp{shara97, gilliland98}).  
This clearly indicates that some process which
increases the initial mass of single stars must be at work. The most
probable formation mechanisms of these peculiar objects are thought to
be either the mass transfer (MT) activity between binary companions,
even up to the complete coalescence of the system~\citep[][]{mcrea64}, or the merger of
two single or binary stars driven by stellar collisions
\citep[COL;][]{zinnsearle76,hillsday76}. Unfortunately, it is still
a quite hard task to distinguish BSSs formed by either channel.  
The most promising route seems to be high resolution
spectroscopy, able to identify chemical anomalies (as a significant
depletion of carbon and oxygen) on the BSS surface
\citep[see][]{fe06_COdep}, as it is expected in the MT scenario. Also
the recent discovery of a double sequence of BSSs in M30
\citep{fe09_m30} seems to indicate that, at least in GCs that recently
suffered the core collapse, COL-BSSs can be distinguished from MT-BSSs
on the basis of their position in the color-magnitude diagram (CMD).

In general, since collisions are most frequent in high density
regions, BSSs in different environments could have different origins:
those in loose GCs might arise from the coalescence/mass-transfer of
primordial binaries, while those in high density clusters might form
mostly from stellar collisions \citep[e.g.,][]{ba92,fe95}. However,
\citet{kn09} suggested that most BSSs, even those observed in the
cores of high density GCs, formed from binary systems, although no
strong correlation between the number or specific frequency of BSSs
and that of binaries is found if all clusters (including the post-core
collapse ones) are taken into account \citep{mil12, leigh13}. This is
likely due to the fact that dynamical processes significantly modify
the binary and the BSS content of GCs during their evolution. An
exception seems to be represented by low density environments, where
the efficiency of dynamical interactions is expected to be
negligible. Very interestingly, indeed, a clear correlation between
the binary and the BSS frequencies has been found in a sample of 13
low density GCs \citep{sol08}.  This remains the cleanest evidence
that the unperturbed evolution of primordial binaries is the dominant
BSS formation process in low-density environments \citep[also
  consistently with the results obtained in open clusters;
  e.g.][]{math09}.

One of the most notable effects of stellar dynamics on BSSs is the
shaping of their radial distribution within the host cluster.  In most
cases a clear bimodality has been observed in the radial distribution
of the $N_{\rm BSS}/N_{\rm ref}$ ratio ($N_{\rm BSS}$ and $N_{\rm
  ref}$ being, respectively, the number of BSSs and the number of
stars belonging to a reference population, as red giant or horizontal
branch stars): such a ratio shows a high peak at the cluster centre, a
dip at intermediate radii and a rising branch in the external regions
\citep[e.g.][]{be08,da09}.  In a few other GCs \citep[see, for
  example, NGC 1904 and M75;][respectively]{lanz07_1904,con12} the
radial distribution shows a clear central peak but no upturn in the
external cluster regions. In the case of $\omega$Centauri
\citep{fe06_ocen}, NGC 2419 \citep{da08} and Palomar 14 \citep{be11}
the BSS radial distribution is flat all over the entire cluster
extension. Since BSSs are more massive than the average cluster stars
(and therefore suffer from the effects of dynamical friction), these
different observed shapes have been interpreted as due to different
``dynamical ages'' of the host clusters \citep[][hereafter F12]{fe12}:
a flat BSS distribution is found in dynamically young GCs
(\emph{Family I}), where two-body relaxation has been ineffective in
  establishing energy-equipartition in a Hubble time and dynamical
  friction has not segregated the BSS population yet. In clusters of
intermediate ages (\emph{Family II}), the innermost BSSs have already sunk
to the center, while the outermost ones are still evolving in isolation
(this would produce the observed bimodality); finally, in the most
evolved GCs (\emph{Family III}), dynamical friction has already segregated
toward the centre even the most remote BSSs, thus erasing the external
rising branch of the distribution. These results demonstrate that the
BSS radial distribution can be used as powerful dynamical clock. Of
course, the same should hold for binary systems with comparable
masses, but the identification of clean and statistically
  significant samples of binaries all over the entire cluster
  extension is a quite hard task and no investigations have been
performed to date to confirm such an expectation.

NGC5466 is a high galactic latitude GC ($l=42.2^{\circ}$ and $b=73.6^{\circ}$), located in the constellation of 
Bo\"{o}tes at a distance of $D=16$ kpc~\citep[][]{ha96}. The cluster has been recently found to be surrounded by huge 
tidal tails~\citep[][]{bel06,gr06}.
The BSS stars in this cluster were first studied by~\citet[][]{ne87}, and, more recently, ~\citet[][]{fek07} analyzed the BSS 
population inside the first $r\sim 800\arcsec$ from the cluster center. Since the cluster tidal radius is $1580\arcsec$~\citep[][]{mio13}, a 
survey of the BSS population over the whole body of the cluster is still missing. In the framework of the study of the origin of 
the BSS, it is important to mention that NGC~5466 is the first GC where eclipsing binaries were found among the 
BSS~\citep[][see also Kryachko et al.(2011)]{ma90}. In their paper~\citet[][]{ma90} find indications that all the BSS in 
NGC 5466 were formed from the merger of the components of primordial, detached binaries evolved into compact binary systems.

In this paper we intend to study the BSS and binary populations of NGC~5466, focussing
in particular on their radial distribution within the whole cluster to investigate its dynamical age.

The paper is organized as follows. In Section \ref{sec_obs} we
describe the observations and data analysis. Section \ref{sec_bss} is
devoted to the definition of the BSS population and the study of its
radial distribution.  In Section \ref{sec_bin} we describe how the
fraction of binaries and its variation with radius are
investigated. The summary and discussion of the obtained results are
presented in Section \ref{sec_concl}.

\section{Observations and data analysis}
\label{sec_obs}
The photometric data used here consist of a combination of shallow and
deep images (see Table~1) acquired through the blue
channel of the Large Binocular Camera (LBC-blue) mounted on the Large
Binocular Telescope (LBT), MEGACAM on the Canadian-France-Hawaii
Telescope (CFHT) and the Advanced Camera for Survey (ACS) on board the
Hubble Space Telescope (HST).

{\it (i)---The LBC data-set}: both short and long exposures were
secured in 2010 under excellent seeing conditions (0\farcs5-0\farcs7)
with the LBC-blue (\citealt{rag06,gial08}) mounted on
the LBT \citep{hill06} at Mount Graham, Arizona.  The LBC is a double
wide-field imager which provides an effective $23' \times 23'$ field
of view (FoV), sampled at $0.224$ arcsec/pixel over four chips of
$2048 \times 4608$ pixels each.  LBC-blue is optimized for the UV-blue
wavelengths, from 320 to 500 nm, and is equipped with the $U-BESSEL$, $B-BESSEL$,
$V-BESSEL$, $g-SLOAN$ and $r-SLOAN$ filters (hereafter $U,B,V,g,r$, respectively).  
A set of short exposures (of 5, 60 and 90
seconds) was secured in the $B$ and $V$ filters with the cluster
center positioned in the central chip (namely \#2) of the LBC-blue CCD
mosaic. Deep images (of 400 and 200 s in the $B$ and $V$ bands,
respectively) were obtained with the LBC-blue FoV positioned
$\sim 100\arcsec$ south from the cluster centre.  A dithering pattern
was adopted in both cases thus to fill the gaps of the CCD
mosaic.  The raw LBC images were corrected for bias and flat field,
and the overscan region was trimmed using a pipeline specifically
developed for LBC image pre-reduction from the LBC Survey data
center\footnote{http://deep01.oa-roma.inaf.it/index.php}.

{\it (ii)---The MEGACAM data-set}: $g$ and $r$ wide-field MEGACAM
images (Prop ID: 04AK03; PI: Jang-Hyun Park), acquired with excellent seeing conditions (0\farcs6-0\farcs8)
were retrieved from the Canadian Astronomy Data Centre
(CADC\footnote{http://www1.cadc-ccda.hia-iha.nrc-cnrc.gc.ca/en/}).
Mounted at CFHT (Hawaii), the wide-field imager MEGACAM (built by CEA,
France), consists of 36 $2048 \times 4612$ pixel CCDs (a total of 340
megapixels), fully covering a 1 degree $\times$ 1 degree FoV with a
resolution of 0.187 arcsecond/pixel.  The data were preprocessed
(removal of the instrumental signature) and calibrated (photometry and
astrometry) by Elixir pipeline\footnote{http://www.cfht.hawaii.edu/Instruments/Elixir/home.html}.

{\it (iii)---The ACS data-set} consists of a set of high-resolution,
deep images obtained with the ACS on board the HST through the F606W
and F814W filters (GO-10775;PI: Sarajedini). The cluster was centered
in the ACS field and observed with one short and five long-exposures
per filter, for a total of two orbits.  We retrieved and used only the
deep exposures (see Table \ref{tab_obs}).  All images were passed
through the standard ACS/WFC reduction pipeline.

\begin{figure} \centering
\includegraphics[width=0.49\textwidth]{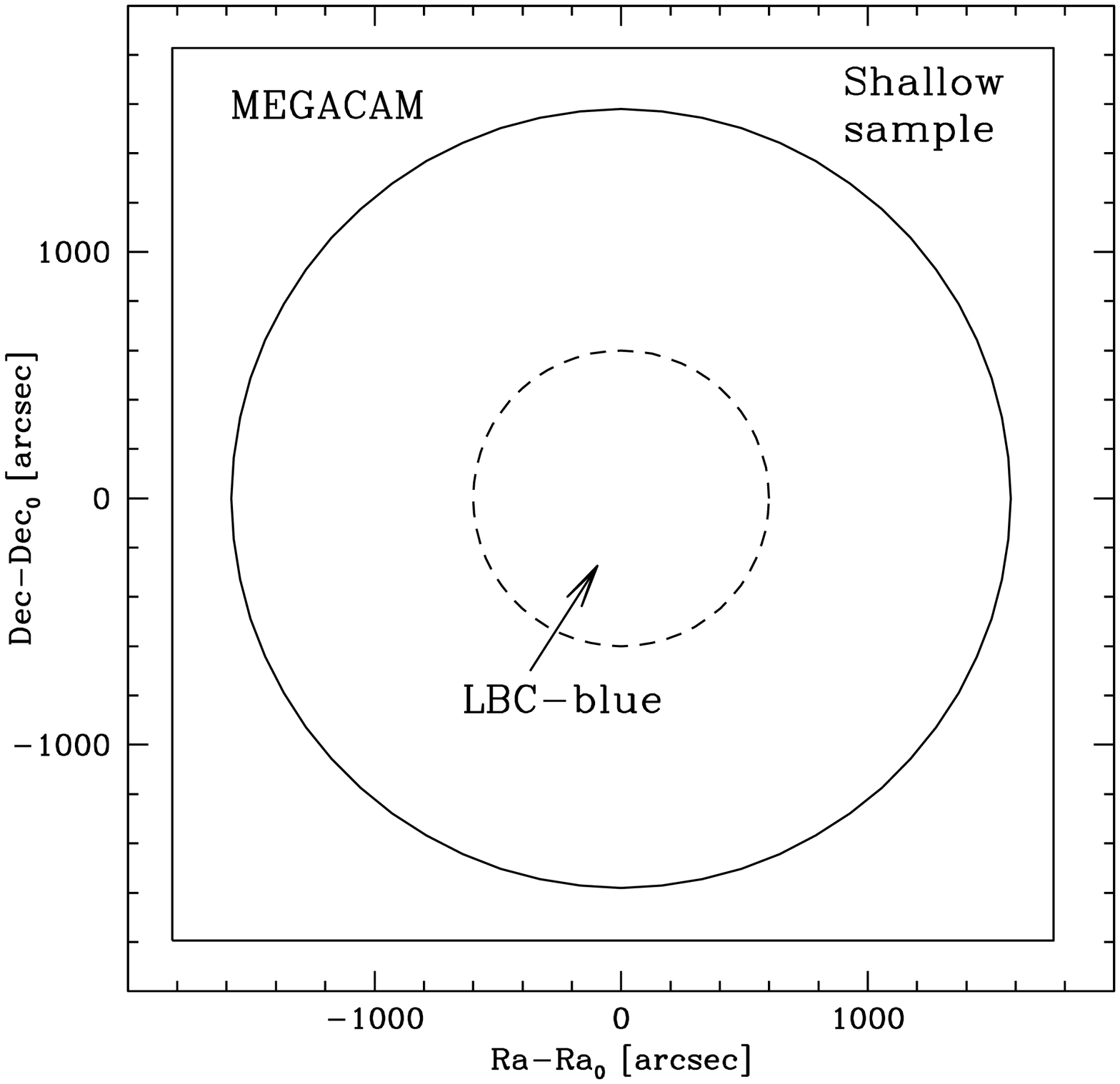}
\includegraphics[width=0.49\textwidth]{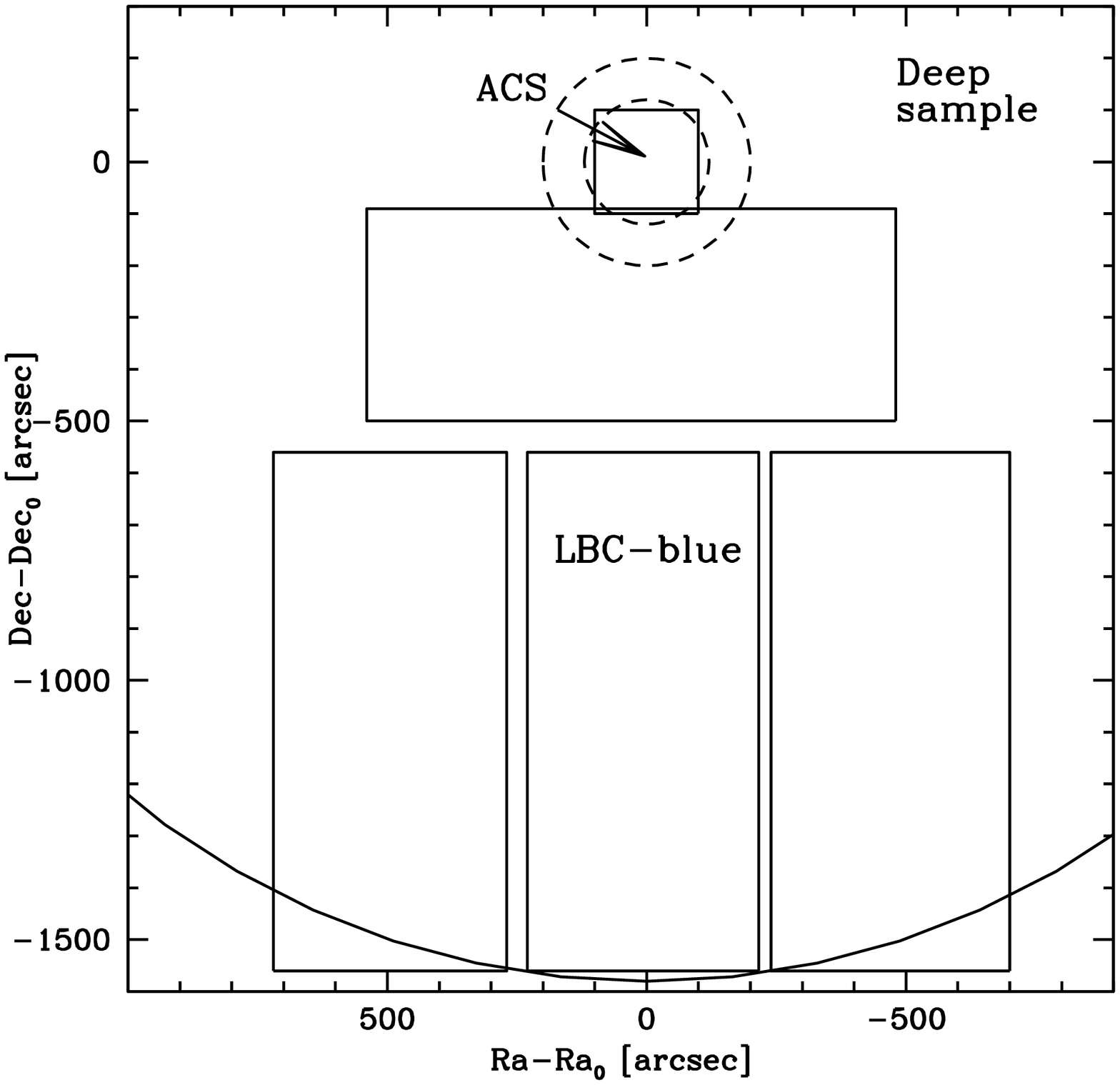}
\caption{{\it Left panel:} map of the ``Shallow Sample'' used to study
  the BSS population of NGC 5466.  The dashed circle defines the
  cluster area probed by the shallow LBC-blue data, while the outer
  square corresponds to the MEGACAM field of view. The solid circle
  corresponds to the location of the cluster's tidal radius.  {\it Right panel:}
  map of the ``Deep Sample'' used to study the cluster binary
  fraction. The solid square corresponds to the ACS FoV, while the
  four rectangles mark the FoV of the deep LBC-blue data-set.  The
  solid circle indicates the location of the cluster's tidal radius. 
  The annulus within the
  two dashed circles shows the area excluded from this study because
  of the low photometric completeness (see Section 4.1).}
\label{maps}
\end{figure}

As shown in the left panel of Figure \ref{maps}, the shallow LBC
data-set is used to sample a region of $600\arcsec$ around the cluster
center, and for $r>600\arcsec$ it is complemented with the MEGACAM
observations, which extend beyond the tidal radius, out to $r\sim
30\arcmin$.  In the following the combination of these two data-sets
will be referred to as the ``Shallow Sample'' and adopted to study the
BSS population over the entire cluster extension (see
Sect. \ref{sec_bss}).  The ACS observations sample the innermost $\sim
200\arcsec$ of the cluster and they are complemented with the deep LBC
observations, extending out to $r_t$ (see right panel of Figure
\ref{maps}): this will be called the ``Deep Sample'' and used to
investigate the binary fraction of NGC 5466 (see Sect. \ref{sec_bin}).

\subsection{Photometry and Astrometry}  
\label{photometry}
Given the very low crowding conditions in the external cluster
regions, we performed aperture photometry on the two $g$ and $r$
MEGACAM images, by using SExtractor \citep{BA96} with an aperture
diameter of $0.9\arcsec$ (corresponding to a FWHM of 5 pixels).  Each of
the 36 chips was reduced separately, and we finally obtained a catalog
listing the relative positions and magnitudes of all the stars in
common between the $g$ and $r$ catalogs.  The LBC and ACS data-sets
were reduced through a standard point spread function (PSF) fitting
procedure, by using DAOPHOTII/ALLSTAR~\citep{st87,st94}. The PSF was
independently modeled on each image using more than 50 isolated and
well sampled stars in the field.  The photometric catalogs of each
pass-band were combined, and the instrumental magnitudes were
reported to a reference frame following the standard procedure
\citep[see, e.g.,][]{fe91,fe92}. The magnitudes of stars successfully
measured in at least three images per filter were then averaged, and
the error about the mean was adopted as the associated photometric
uncertainty.

In order to obtain an absolute astrometric solution for each of the 36
MEGACAM chips, we used more than 15000 stars in common with the 
Sloan Digital Sky Survey (SDSS)
catalog\footnote{Available at web-site
  http://cas.sdss.org/dr6/en/tools/search/radial.asp} covering an area
of 1 square degree centered on the cluster. The final r.m.s. global
accuracy is of $0.3\arcsec$, both in right ascension ($\alpha$) and
declination ($\delta$).  The same technique applied to the LBC sample
gave an astrometric solution with similar accuracy
($\sigma<0.3\arcsec$ r.m.s.). Considering that no standard astrometric
stars can be found in the very central regions of the cluster, we used
the stars in the LBC catalog as {\it secondary astrometric standards}
and we thus determined an absolute astrometric solution for the ACS
sample in the core region with the same accuracy obtained in the
previous cases.
 
A sample of bright isolated stars in the ACS data-set was used to
transform the ACS instrumental magnitudes to a fixed aperture of
$0\farcs5$. The extrapolation to infinite and the transformation into
the VEGAMAG photometric system was then performed using the updated
values listed in Tables 5 and 10 of \citet[][]{sir05}\footnote{The new
values are available at the STScI web pages:
http://www.stsci.edu/hst/acs/analysis/zeropoints}. In order to transform the instrumental $B$ and $V$ magnitudes of the LBC sample
into the Johnson/Kron-Cousins standard system, we used the stars in common with a
photometric catalog previously published by \citet{fek07}. 
These authors, in section 2.3 of their paper, provide a thorough comparison of their
photometry with existing literature datasets, finding a satisfactory agreement.
In particular they found an excellent agreement
in the photometric zero-points (within less than 0.02 mag) and no residual trends with colors
with the set of secondary standards by~\citet[][]{ste00}, that should be considered as
the most reliable reference. We verified, by direct comparison with Stetson (2000), that the 
quality of the agreement shown by \citet{fek07} is fully preserved in our final catalog.

Instead, the $g$ and $r$ magnitudes in the MEGACAM sample were calibrated using
the stars in common with the SDSS catalog and we then adopted the
equation that Robert Lupton derived by matching DR4 photometry to
Peter Stetson's published photometry for stars\footnote{see
  http://www.sdss.org/dr4/algorithms/sdssUBVRITransform.html\#Lupton2005}
to transform the calibrated $g$ and $r$ into a $V$ magnitude.  We
finally used more than 2000 stars in common between the LBC and the
MEGACAM data-sets to search for any possible off-sets between the $V$
magnitude in the two catalogs, finding an average systematic
difference of 0.014 mag.  We therefore applied this small correction
to the MEGACAM catalog.  Similarly, we used more than 200 stars in
common between the ACS data-set and the LBC catalog to calibrate the
ACS F606W magnitude to the Johnson $V$ magnitude.  Such a procedure
therefore provided us with a homogenous $V$ magnitude scale in common
to all the available data-sets.

\subsection{The Color-Magnitude Diagrams}
\label{sec_cmd}
\begin{figure}
\plotone{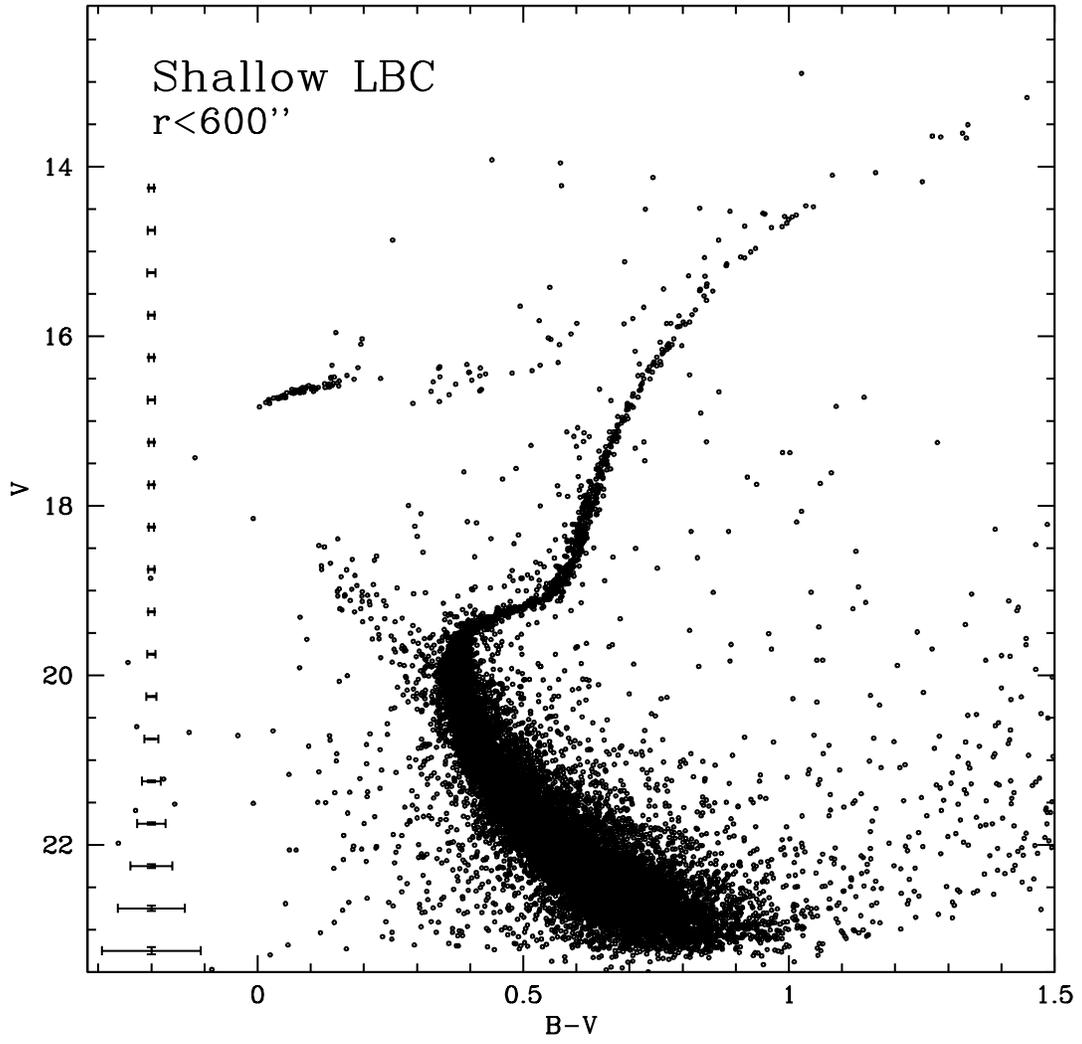} 
\caption{CMD of the stars sampled with the shallow LBC data-set in an
  area of $10\arcmin$ around the cluster's center.}
\label{cmd_shallow}
\end{figure}

The CMD obtained from the shallow LBC exposures for the innermost
$600\arcsec$ from the cluster center is shown in
Figure~\ref{cmd_shallow}. Typical photometric errors (magnitudes and colors) are indicated by black crosses. 
The high quality of the images allows us to
accurately sample all the bright evolutionary sequences, from the Tip
of the red giant branch (RGB), down to $\sim 2$ magnitudes below the
MS-TO, with a large population of BSSs well visible at $0<(B-V)<0.4$
and $20.5<V<18$.

\begin{figure}
\plotone{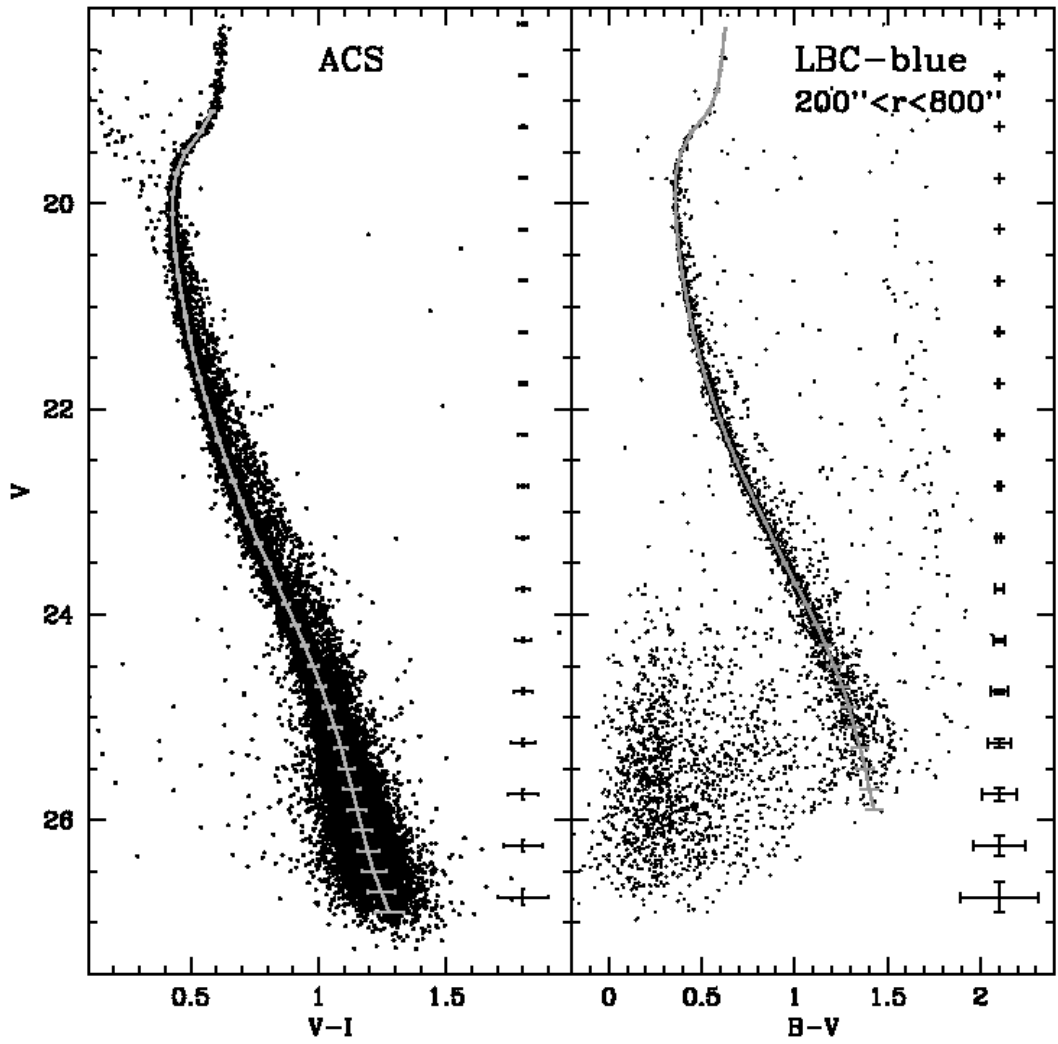} 
\caption{CMDs of the ACS and the deep LBC samples (left and right
  panels, respectively).  The mean ridge line is shown as a solid
  line.} 
\label{cmd_deep}
\end{figure}

In Figure \ref{cmd_deep} we show the CMDs obtained from the ACS and
the deep LBC observations. As in Figure~\ref{cmd_shallow}, the typical photometric errors (magnitudes and colors) 
for the two sample are indicated by black crosses.
 A well populated MS is visible in the ACS
CMD, extending from the MS-TO ($V\sim20.5$) down to $V\sim27$.  Once
again, the collecting power of the LBT combined with the LBC-blue
imaging capabilities allowed us to sample the MS down to similar
magnitudes in an area extending out to the cluster's tidal radius.
For $(B-V)\gtrsim 1.5$ the plume of Galactic M dwarfs is clearly
visible, while for $(B-V)\lesssim 1$ and $V\gtrsim 24$ a population of
unresolved background (possibly extended) objects is apparent.  In
both the CMDs a well defined sequence of unresolved binary systems is
observed on the red side of the MS. 

%{\it , which
%  location is defined trough the fit with a 12 Gyr model
%  from~\citet[][dashed grey line in the figure]{dot07}. The best fit
%  model was obtained using an isochrone at the metallicity [Fe/H] =
%  -2.22, distance moduli $(m - M)_V=16.16$ and adopting a reddening,
%  $E(B-V) = 0.0$ \citep[][]{f99}.}

\section{The shallow sample and the BSS population}
\label{sec_bss}
The Shallow Sample was used to determine the cluster center of
gravity ($C_{\rm grav}$) and the projected density profile from
accurate star counts\footnote{The observed profile is shown in \citet[][]{mio13}, 
as part of a catalog of 26
Galactic GCs for which also the best-fit~\citet[][]{k66} and~\citet[][]{w75} model fits
are discussed.}, following the procedure already adopted and
described, e.g., in \citet{be12}.  The resulting value of $C_{\rm grav}$ 
is $\alpha_{\rm J2000} =14^{\rm h}\,05^{\rm m}\, 27^{\rm s}.2$, 
$\delta_{\rm J2000} =+28^\circ\,32^\prime\, 01\farcs8$, in
full agreement with that quoted by \citet{g10}.  The best fit King
model to the observed density profile has concentration parameter
$c=1.31$, core radius $r_c=72\arcsec$ and tidal radius
$r_t=1580\arcsec =26.3\arcmin$~\citep[][]{mio13}.\footnote{Here we
  call ``tidal radius'' what in~\citet[][]{mio13} is named `limiting
  radius'' ($r_\ell$). While the latter is formally more correct, we
  keep referring to the tidal radius for sake of continuity with our
  previous papers about the BSS populations in GCs (see F12 and
  references therein).}  

\citet{fek07} studied the radial distribution of the BSS population in
NGC 5466, from the cluster centre out to $r\sim 800\arcsec$ \citep[see
  also][]{ne87}.  They  selected 89 BSSs and showed that they are
significantly more centrally concentrated than RGB and horizontal
branch (HB) stars, but they did not find any evidence of upturn in the
external regions.  However, that study did not sample the entire
radial extension of the cluster, while our previous experience
demonstrates that even a few BSSs in the outer regions can generate
the external rising branch. Hence, a detailed investigation even in
the most remote regions is necessary before solidly excluding the
presence of an upturn in the BSS radial distribution and confirming
the dynamical age of the cluster.

\subsection{BSS and reference population selection}
\label{sec_sampl}
\begin{figure}
\plotone{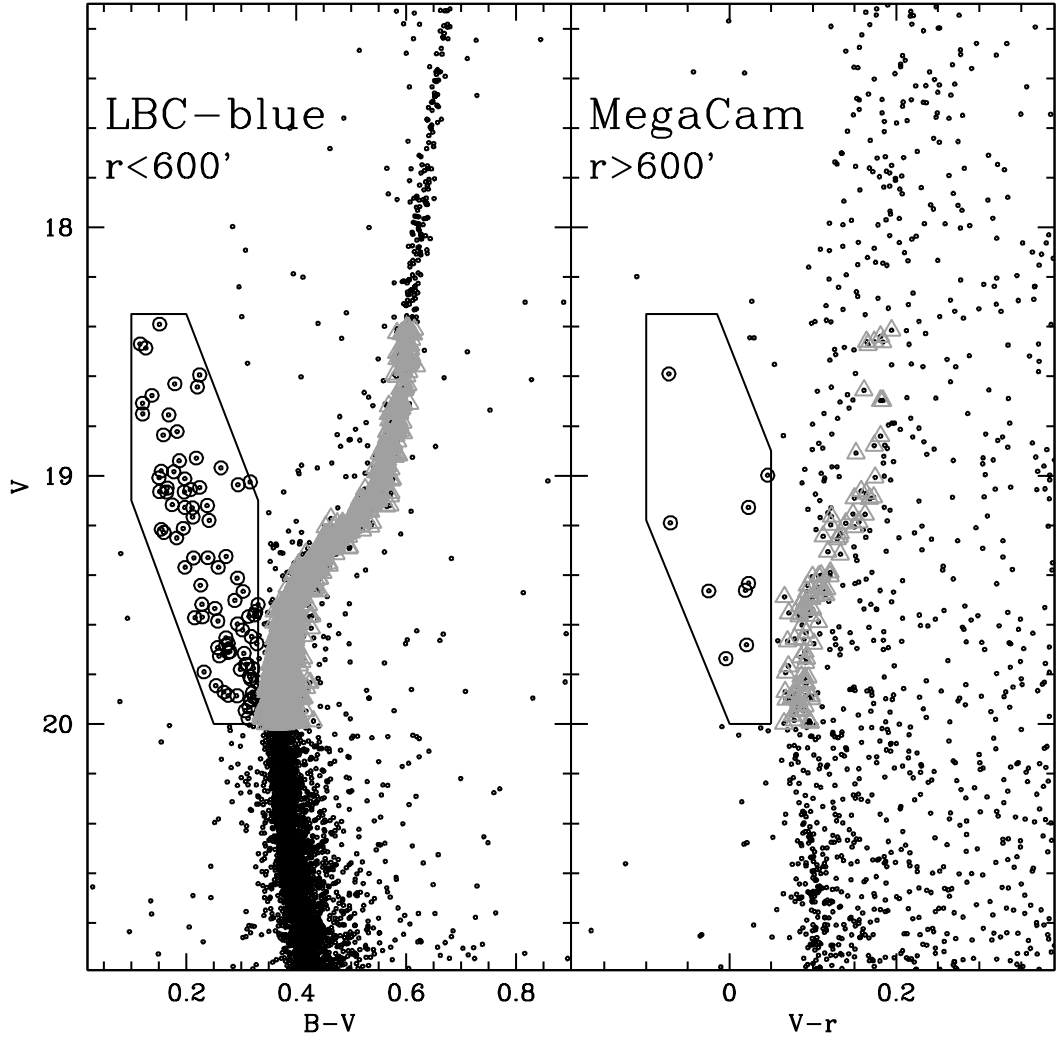} 
\caption{CMDs showing the selection of the BSS population (open
  circles) in the shallow LBC and the MEGACAM samples (left and right
  panels, respectively). The grey open triangles marks the population
  of sub-giant and RGB stars used as reference in the study of the BSS
  radial distribution.}
\label{cmd_bss}
\end{figure}

The selection of the BSS population was primarily performed in the
$(V,B-V)$ plane (see open circles in Figure \ref{cmd_bss}, left
panel), by adopting conservative limits in magnitude and color in
order to avoid the possible contamination from stellar blends.  The
BSS selection has been then ``exported'' to the MEGACAM $(V,V-r)$ CMD
by using the BSSs in common between the two data-sets in the region at
$r<600\arcsec$ (see right hand panel).  We finally count a population
of 88 stars in the LBC sample and 9 stars in the MEGACAM one, for a
total of 97 BSSs spanning the entire radial extent of the cluster.
The cross-correlation of our sample with that of \citet{fek07}
confirms that the majority of the stars selected in the previous work
are real BSSs, but 18 of them are blended sources. In order to
  further test the completeness of our LBC shallow sample we performed
  a detailed comparison with the ACS photometric catalogue of
  \citet{sa07}\footnote{The catalogue is available for download at
    http://www.astro.ufl.edu/$\sim$ata/public\_hstgc/}.  The completeness
  has been quantified as the fraction of stars in common between the
  two catalogs in a given magnitude interval: we find 93\%, 95\%, and
  100\% completness for stars with $V< 22$, 21.5, and 21,
  respectively.

For a meaningful study of the BSS population, a reference sample of
``normal'' cluster stars has to be defined. Here we choose a post-MS
population composed of sub-giant (SGB) and red giant branch stars (RGB; see
grey triangles in Fig. \ref{cmd_bss}). This has been selected in the
same magnitude range ($18.4<V<20$) of BSSs, thus guaranteeing that the
photometric completeness is the same in both samples. We count 1358
and 96 stars in the LBC and MEGACAM data-sets, respectively.

\subsection{The BSS radial distribution}
\label{sec_bssrad}
As extensively illustrated in literature \citep[see for example][and
  references therein]{ba95,fe03, fe12}, the study of the BSS radial
distribution is a very powerful tool to shed light on the internal
dynamical evolution of GCs and possibly gets some hints on the BSS
formation mechanisms.

\begin{figure}
\plotone{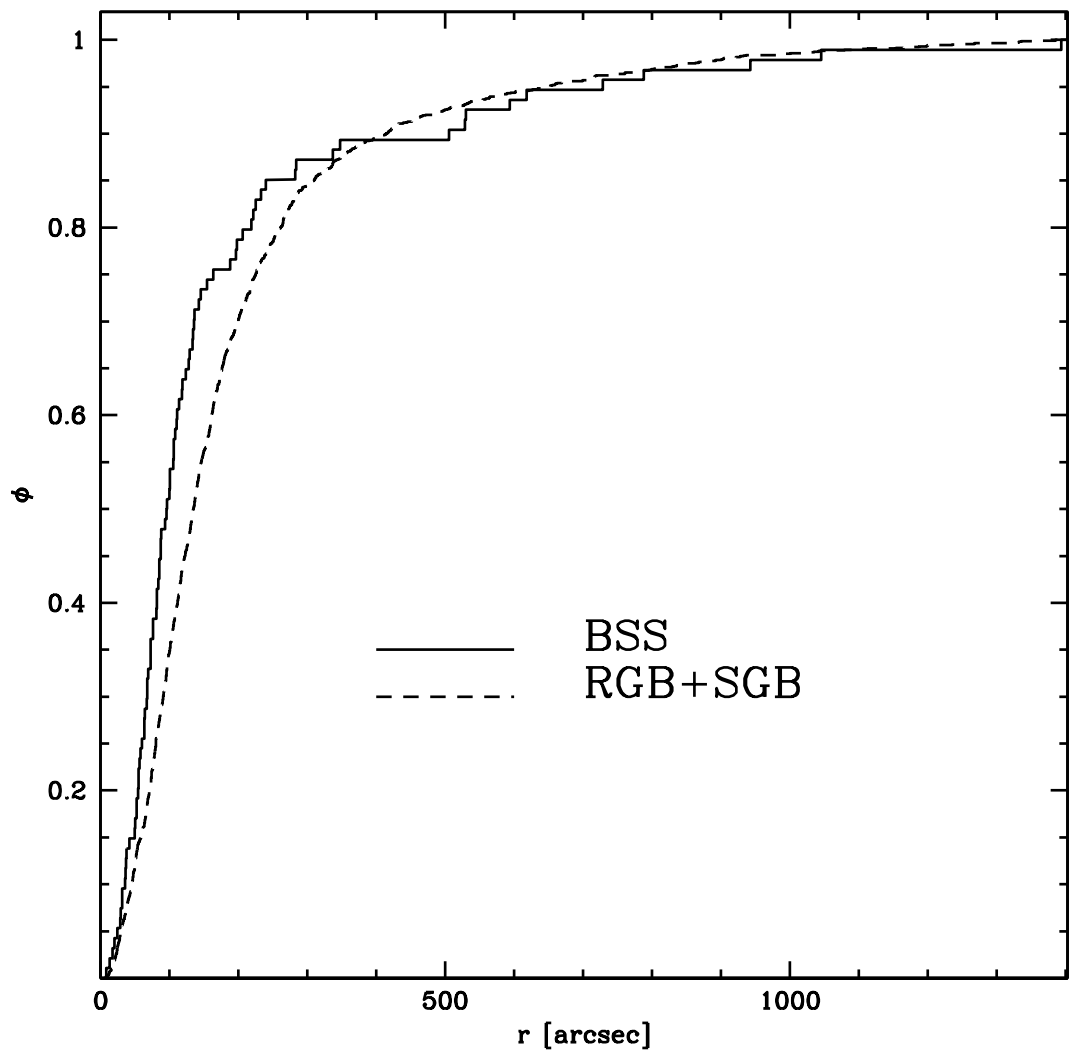} 
\caption{Cumulative radial distribution of BSSs (solid line) and
  sub-giant+RGB stars (dashed line), as a function of the projected
  distance from the cluster center.}
\label{ks}
\end{figure}

As a first step, we compared the cumulative radial
distribution of the selected BSS population to that of reference stars
(see Figure \ref{ks}).  Clearly, BSSs are more centrally
concentrated, and a Kolmogorov-Smirnov test yields a probability of
$99.9\%$ that the two samples are not extracted from the same parent
population. This cumulative BSS distribution closely resembles that
already observed in this cluster by \citet{fek07} out to $\sim
800\arcsec$ (see their Figure 19), with the only difference that we
sample the entire cluster extension.  Moreover, the distributions
shown in Figure \ref{ks}, closely resembles that observed in all
GCs characterised by a bimodal radial distribution of BSSs (see, e.g.,
the cases of M3, \citealp{f97}, 47Tuc, \citealp{f04}; M53,
\citealp{r98}; M5, \citealp{lanz07_M5}).

Following the same procedure described in \citet{lanz07_M5}, as a second
step we divided the surveyed area in a series of concentric annuli
centred on $C_{\rm grav}$. In Table~2 we report the values
of the inner and outer radii of each annulus, together with the number
of BSSs ($N_{\rm BSS}$) and reference stars ($N_{\rm RGB}$) counted in
each radial bin. We also report the fraction of light sampled in the
same area, with the luminosity being calculated as the sum of the
luminosities of all the stars measured in the TO region at
$19.5\leqslant V \leqslant20.5$.  The contamination of the cluster BSS
and RGB populations due to background and foreground field stars can
be severe, especially in the external regions where number counts are
low. We therefore exploited the wide area covered by MEGACAM (see
Figure \ref{maps}) to estimate the field contamination by simply
counting the number of stars located beyond the cluster tidal radius
(at $r>1580\arcsec$) that populate the BSS and RGB selection boxes in the CMD.
We count 2 BSSs and 13 RGB stars in a $\sim0.43$ deg$^2$ area.  The
corresponding densities are used to calculate the number of field
stars expected to contaminate each population in every considered
radial bin (see numbers in parenthesis in Table~2).

In the upper panel of Figure \ref{bss_rad} we show the resulting trend
of $N_{\rm BSS}/N_{\rm RGB}$ as a function of radius. 
%% Taking advantage of the wide field capabilities of MEGACAM we are
%% able to extend the radial distribution to the cluser's $t_r$. ****
Clearly it is bimodal, with a relatively low peak of BSSs in the inner
region, a distinct dip at intermediate radii, and a constant value
slightly smaller than the central one, in the outskirts. We identify
the minimum of the radial distribution at $r_{\rm min}\sim180\arcsec$.
This behavior is further confirmed by the radial distribution of the
BSS double normalized ratio, $R_{\rm BSS}$ (see the black circles in
the lower panel of Figure \ref{bss_rad}).

For each radial bin, $R_{\rm BSS}$ is defined
as the ratio between the fraction of BSSs in the annulus and the
fraction of light sampled by the observations in the same bin
\citep[see][]{fe03}.  The same quantity computed for the reference
population turns out to be equal to one at any radial distance (see
grey regions in the figure). This is indeed expected from the stellar
evolution theory \citep{ren89} and it confirms that the distribution
of RGB stars follows that of the cluster sampled light. Instead,
$R_{\rm BSS}$ shows a bimodal behavior, confirming the result
obtained above. Notice that NGC 5466 is one of the most sparsely
populated Galactic GCs, and so the number of cluster's stars and the
fraction of sampled light in the external regions is quite low. 
%As shown in Table~2, only 2\% of the cluster light is
%included in the last annulus.

\section{The deep sample and the cluster binary fraction}
\label{sec_bin}
Binaries play an important role in the formation and dynamical
evolution of GCs~\citep[][]{mc91}. Indeed these systems secure a enormous energy
reservoir able to quench mass-segregation and delay or even prevent
the collapse of the core.  Up to now three main
techniques have been used to measure the binary fraction: {\it (i)}
radial velocity variability \citep[e.g.,][]{math09}, {\it (ii)}
searches for eclipsing binaries \citep[e.g.][]{mat96,co96} and {\it
  (iii)} studies of the distribution of stars along the cluster MS in
CMD \citep[e.g.][]{ro91,be02,mil12}.  In this work we adopt the latter
approach, which has the advantage of being more efficient for a
statistical investigation, and unbiased against the orbital parameters
of the systems.  We followed the prescriptions extensively described
in~\citet[][]{be02} and~\citet[][see also Dalessandro et
  al. 2011]{sol07, sol10}.  This method relies on the fact that the luminosity of a binary
system is the result of the contributions of both (unresolved)
companions.  The total luminosity is therefore given by the luminosity
of the primary (having mass $M_1$) increased by that of the companion
(with mass $M_2$).  Along the MS, where stars obey a mass-luminosity
relation, the magnitude increase depends on the mass ratio
$q=M_2/M_1$, reaching a maximum value of 0.75 mag for equal mass
binaries ($q=1$). Taking advantage of the large FoV covered by our
data-sets, we analyzed the binary fraction at different radial
distances from the cluster center, out to $r\sim800\arcsec$.

\subsection{Completeness tests}

\label{sec_compl}
In order to estimate the fraction of binaries from the ``secondary'' MS
in the CMD, it is crucial to have a realistic and robust measure of
other possible sources of broadening of the MS (e.g. blending,
photometric errors).  These factors are related to the quality of the
data and can be properly studied through artificial star
experiments. We therefore produced a catalog of artificial stars for
the ACS and LBC-deep data-sets, following the procedure described
in~\citet[][]{be02}.  

As a first step, a list of input positions and magnitudes of artificial stars
(X$_{in}$,Y$_{in}$,$V_{in}, F814W_{in}$ and $B_{in}$) is produced. 
Once these stars are added to the original frames at the X$_{in}$ and Y$_{in}$ positions, 
we repeat the measure of stellar magnitudes in the same way as described in Section~\ref{photometry}.   
At the end of this procedure we obtain a list of stars including both real and artificial stars. Since
we know precisely the positions and magnitudes of the artificial stars,
the comparison of the latter with the measured $V_{out}, F814W_{out}$ and $B_{out}$ magnitudes,
allow us to estimate the capability of our data-reduction strategy
to properly detect stars in the images, including the effect of blending.

In practice, we first generated a catalog of
simulated stars with a $V_{in}$ magnitude randomly extracted from a
luminosity function modeled to reproduce the observed one. Then the
$F814W_{in}$ and $B_{in}$ magnitudes were assigned to each sampled $V_{in}$ magnitude
by interpolating the mean ridge line of the cluster (solid grey lines
in the left and right panels of Figure \ref{cmd_deep}, for the ACS and
the LBC-deep samples, respectively).  The artificial stars were then
added on each frame using the same PSF model calculated during the
data reduction phase, and were spatially distributed on a grid of
cells of fixed width (5 times larger than the mean FWHM of the stars
in the frames). Only one star was randomly placed in such a box in
each run, in order not to generate artificial stellar crowding from
the simulated stars on the images. The artificial stars were added to
the real images using the DAOPHOT/ADDSTAR routine. At the end of this
step, we have a number of ``artificial'' frames (which include the real stars and
the artificial stars with know position in the frame) equal to the number of observed frames.

The reduction process was repeated on the artificial images in exactly the same way
as for the scientific ones.  A total of 100,000 stars were simulated
on each data-set and photometric completeness ($\phi_{\rm comp}$) 
was then calculated as the
ratio between the number of stars recovered by the photometric
reduction ($N_{\rm out}$) and the number of simulated 
stars ($N_{\rm in}$).

In Figure \ref{compl}, we show the photometric completeness estimated
for the ACS and the LBC-deep data-sets (upper and lower panels,
respectively) as a function of the $V$ magnitude and for different
radial regions around the cluster center.  The completeness of the ACS
sample is above 80\% for all magnitudes down to $V\sim 26$ (i.e.,
$\sim 5.5.$ magnitudes below the cluster TO), even in the innermost
region.  On the other hand, a completeness above 50\% down to $V\sim
24$ is guaranteed only at $r>200\arcsec$ in the LBC-deep data-set.
Instead, stellar crowding is so strong at lower radial distances
(region D) that the completeness stays below $\sim 80\%$ at all
magnitudes. We therefore exclude this region
($120\arcsec<r<200\arcsec$) from our study of the cluster binary
fraction.

\subsection{The binary fractions}
\label{sec_binfrac}

As mentioned at the beginning of Section~\ref{sec_bin}, our estimation of the binary
fraction in NGC~5466 is performed by determining the fraction of binaries
populating the area between the MS and a $secondary$ MS, blue-ward shifted in colour
by $\sim0.75$~\citep[][]{hu98}.

As a first step of our analysis, we measured the minimum binary
fraction, i.e. the fraction of binaries with mass ratios large enough
($q>q_{\rm min}$) that they can be clearly distinguished from single
MS stars in the CMD. When $q$ approaches the null value, the shift in magnitude of
the primary star (i.e. the more massive one) is negligible. For this reason binary systems with 
small values of $q$ becomes indistinguishable from MS stars when photometric errors are present.
Clearly, this is a lower limit to the total binary
fraction of the cluster.  However, it has the advantage of being a pure observational quantity, 
since it is calculated without any  assumption on the distribution of 
mass ratios~\citep{sol07,da11}.
%Of course the value of $q_{\rm min}$ depends on the photometric
%accuracy in the magnitude interval in which the analysis is performed.

The magnitude interval in which the minimum binary fraction has been
estimated is $21<V<24$ for both the ACS and the LBC samples. According
to a 12 Gyr model from~\citet[][]{dot07}\footnote{The best fit model was obtained 
using an isochrone of metallicity [Fe/H] =-2.22, distance modulus 
$(m - M)_V=16.16$ and adopting a reddening,
$E(B-V) = 0.0$ \citep[][]{f99}.}, this corresponds to a mass range
$0.5M_\odot \lesssim M\lesssim 0.74 M_\odot$ for a single star on the
MS.  In this interval the photometric completeness $\phi_{comp}$ is larger
than $50\%$ for both samples.  For the magnitude and
radial intervals just defined, we divided the entire FoV in four
annuli (see Table~3 for details).  

Then, the location of $q_{\rm min}$ was estimated on the CMD
as the value corresponding to a color difference of
three times the photometric error from the MS ridge mean line (grey
solid lines in the cluster's CMDs shown in Figure \ref{cmd_deep}). 
This approach allows us to define an area on the CMD safe from contamination from
single mass stars. We notice that 
for each of the four annuli, this color difference corresponds to $q_{\rm min}=0.5$.
The contamination from blended sources was calculated on the same
region of the CMD by adopting the catalogue of artificial stars 
produced with the completeness tests (see Section~\ref{sec_compl}).
A catalogue of stars from the Galaxymodel of~\citet[][]{rob03} was used to
estimate the field contamination. We thus estimated $\xi_{q\ge 0.5}$
independently for the four radial intervals  
using the same approach described in \citet{sol07}.
As shown in Figure \ref{bin} (upper
panel), we find that $\xi_{q\ge 0.5}$ is almost constant and equal to
$\sim 6.5\%$ for $r\lesssim 90\arcsec$ (i.e., within the ACS sample),
then it drops at $\sim 5\%$ at distances larger than $200\arcsec$ (in
the LBC sample).

As described above, the minimum binary fraction has the advantage of
being measured with no arbitrary assumptions about the mass ratio
distribution $f(q)$. It depends, however, on the photometric accuracy of
the used data-sets, and caution must be exercised when comparing the
minimum binary fractions estimated for different GCs and in different
works.  An alternative approach consists in measuring the global fraction of
binaries ($\xi_{|rm TOT}$), that can be obtained by simulating a
binary population which follows a given distribution $f(q)$, and by
comparing the resulting synthetic CMD with the observed one.  We
already adopted this technique in~\citet{be02}, \citet{sol07} and
\citet{da11}. Hence, we refer to these papers for details.  In brief,
in order to simulate the binary population we randomly extracted
$N_{\rm bin}$ values of the mass of the primary component from a
\citet{kr02} initial mass function, and $N_{\rm bin}$ values of the
binary mass ratio from the $f(q)$ distribution observed by
\citet{fi05} in the solar neighborhood, thus also obtaining the mass
of the secondary. We considered the same magnitude limits and radial
intervals used before.  For each adopted value of the input binary
fraction $\xi_{\rm in}$ (assumed to range between 0 and 15\% with a
step of 1\%), we produced 100 synthetic CMDs.  Then, for each value of
$\xi_{\rm in}$ the penalty function $\chi^2(\xi_{\rm in})$ was
computed and a probability $P(\xi_{\rm in})$ was associated to it
\citep[see][]{sol07}.  The mean of the best fitting Gaussian is the
global binary fraction $\xi_{\rm TOT}$. The values obtained in the
four considered radial bins are listed in Table~3. The
radial distribution of $\xi_{\rm TOT}$ well resembles that obtained
for the minimum binary fraction, remaining flat at $\sim 8\%$ in the
innermost $\sim 90\arcsec$ and decreasing to $\sim 6\%$ in the two
outermost annuli.
 
\section{Summary and Discussion}
\label{sec_concl}
We have investigated the radial distribution of BSSs and binary systems
in the Galactic GC NGC 5466, sampling the entire cluster
extension. This has been possible thanks to a proper combination of
LBC-blue and MEGACAM observations, and deep ACS and LBC-blue data,
respectively.

\begin{figure}
\plotone{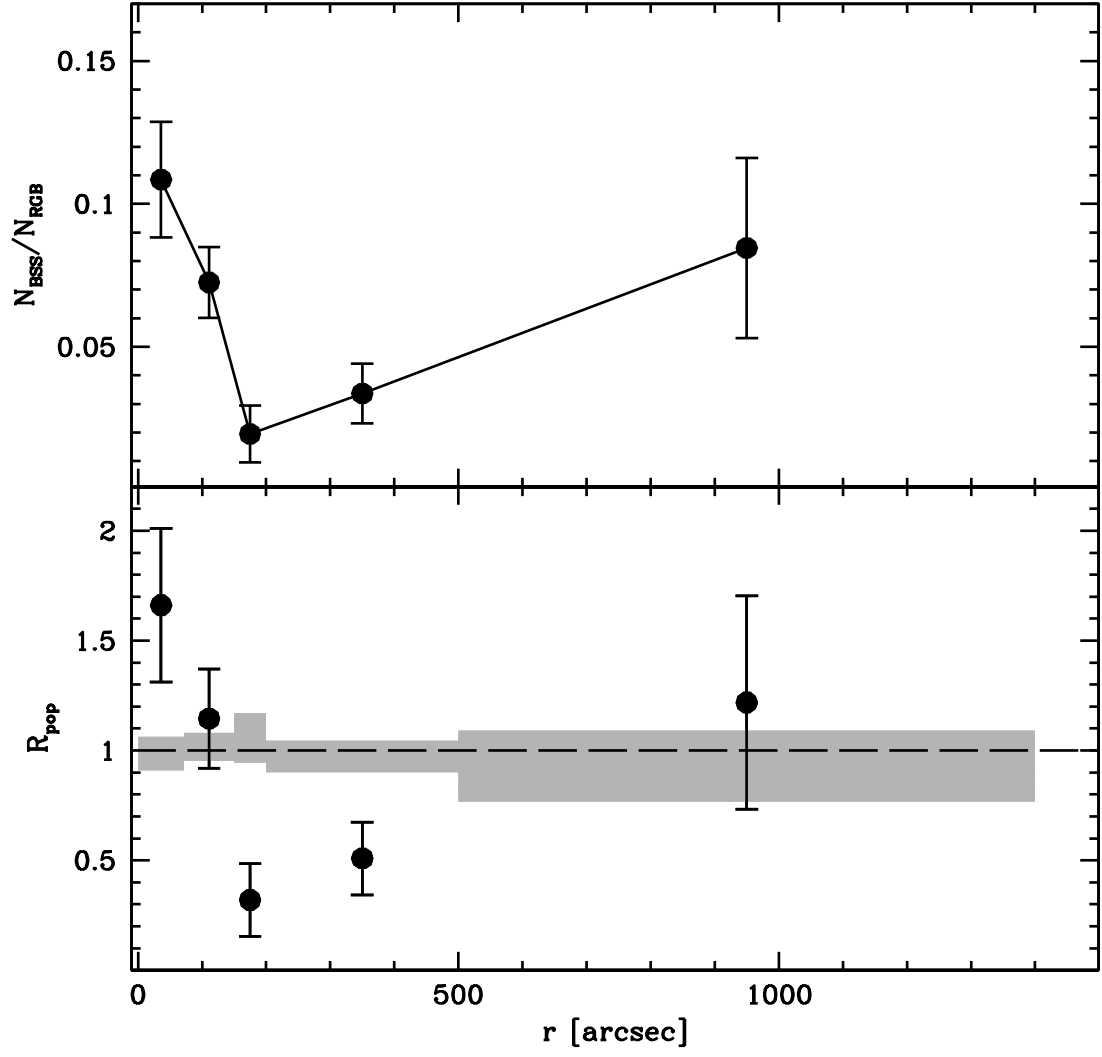} 
\caption{BSS rardial distribution. {\it Upper panel}: number of BSSs
  with respect to that of reference (SGB+RGB) stars, plotted as a
  function of the distance from the cluster center. {\it Lower panel}
  double normalized ratio (see text) of BSSs (dots and error bars) and
  reference stars (gray regions). The distribution is clearly bimodal,
  with a minimum at $r_{\rm min}\simeq180\arcsec$.}
\label{bss_rad}
\end{figure}

\begin{figure}
\plotone{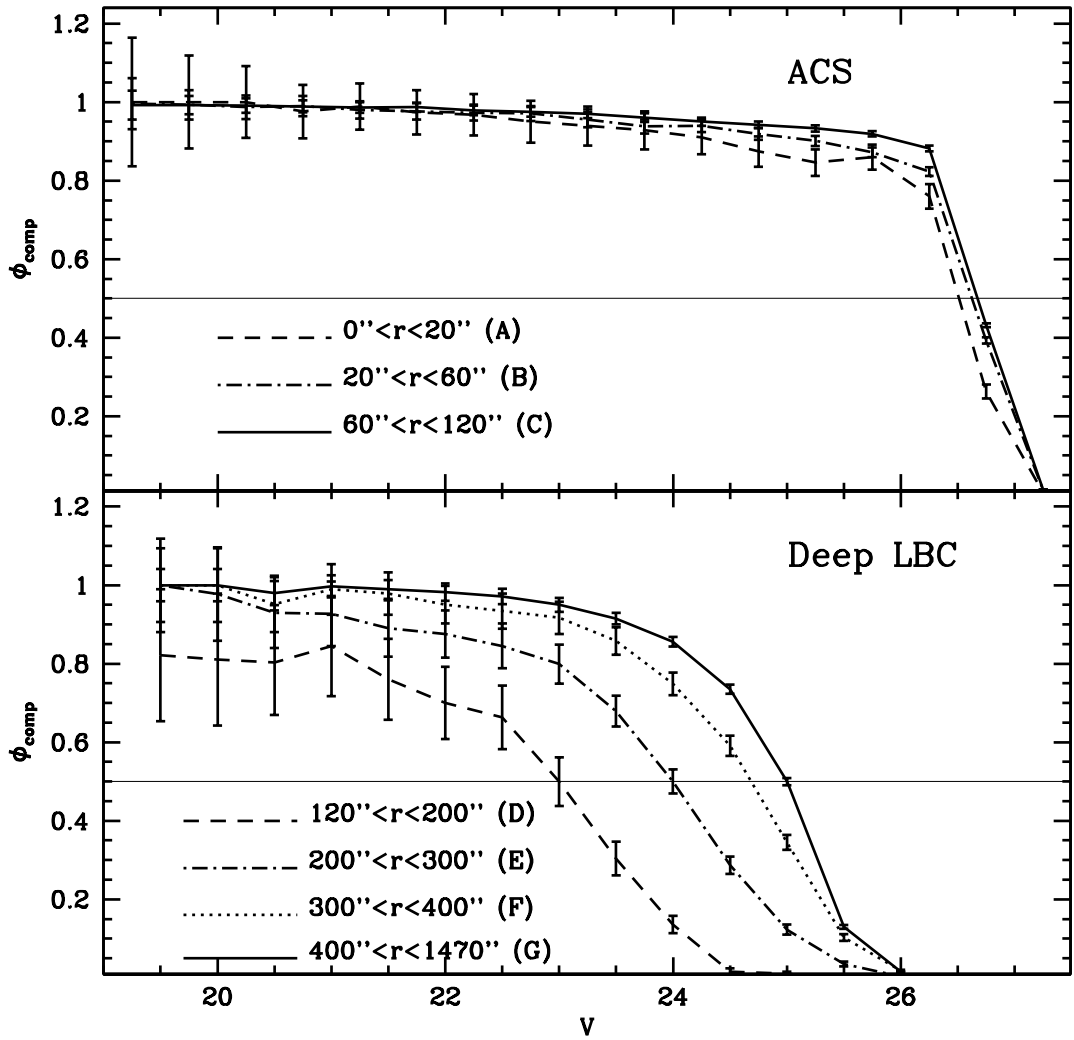} 
\caption{Photometric completeness $\phi$ as a function of $V$
  for the two data sets. The ACS and deep LBC data-sets have been
  divided into three and four concentric annuli areas of same
  completeness, respectively. The solid horizontal line shows the
  limits of 50\% of completeness.}
\label{compl}
\end{figure}

The BSS radial distribution has been found to be bimodal, as in the
majority of GCs investigated to date (see F12 and references therein).
In fact, our data-set allowed us to detect a ``rising branch'' in the
external portion of the BSS distribution that was not identified in
previous investigations because of a too limited sample in radius
\citep[$r<800\arcsec$][]{fek07}.  Following the F12 classification,
NGC 5466 therefore belongs to \emph{Family II}.  The minimum of the
distribution is a quite prominent, purely observational feature, that,
it in the light of the results published by F12, can be used to
investigate the dynamical age of NGC 5466.  It is located at $r_{\rm
  min}\sim 180\arcsec$, corresponding to only $\sim 2.5 r_c$.  This is
the second most internal location of the minimum, after that of M53. Hence, NGC 5466
is one of the dynamically youngest clusters of \emph{Family II}, where
BSSs have just started to sink toward the center of the system.  The
relatively low central peak of the BSS distribution ($R_{\rm
  BSS}\simeq 1.7$; see Fig. \ref{bss_rad}) is also consistent with
this interpretation. As a further check, we calculated the core
relaxation time ($t_{\rm rc}$). F12 shows that $log(t_{\rm rc})$ is proportional
to $log(r_{\rm min})$ (see their figure 4).  The value of $t_{\rm rc}$ has been
computed by using equation (10) of \citet{dj93}, adopting the new
cluster structural parameters reported in Section \ref{sec_cmd}.  In
Figure \ref{clock} we show the position of NGC 5466 (black solid
circle) on the ``dynamical clock plane'' ($t_{\rm rc}/t_{\rm H}$ as a
function of $r_{\rm min}/r_c$; Figure 4 in F12), where $t_{\rm H}$ is
the age of the Universe ($t_{\rm H}$ = 13.7 Gyr).  Clearly, this
cluster nicely follows the same relation defined by the sample
analyzed in F12 (see also \citealp{dale13}), further confirming that
the shape of the observed BSS distribution is a good measure of GC
dynamical ages. This figure also confirms that NGC 5466 is one of the
youngest cluster of \emph{Family II}, meaning that only recently the
action of dynamical friction started to segregate BSSs (and primordial binary
systems of similar total mass) toward the cluster centre. In this
scenario, the most remote BSSs are thought to be still evolving in
isolation in the outer cluster regions.

\begin{figure}
\plotone{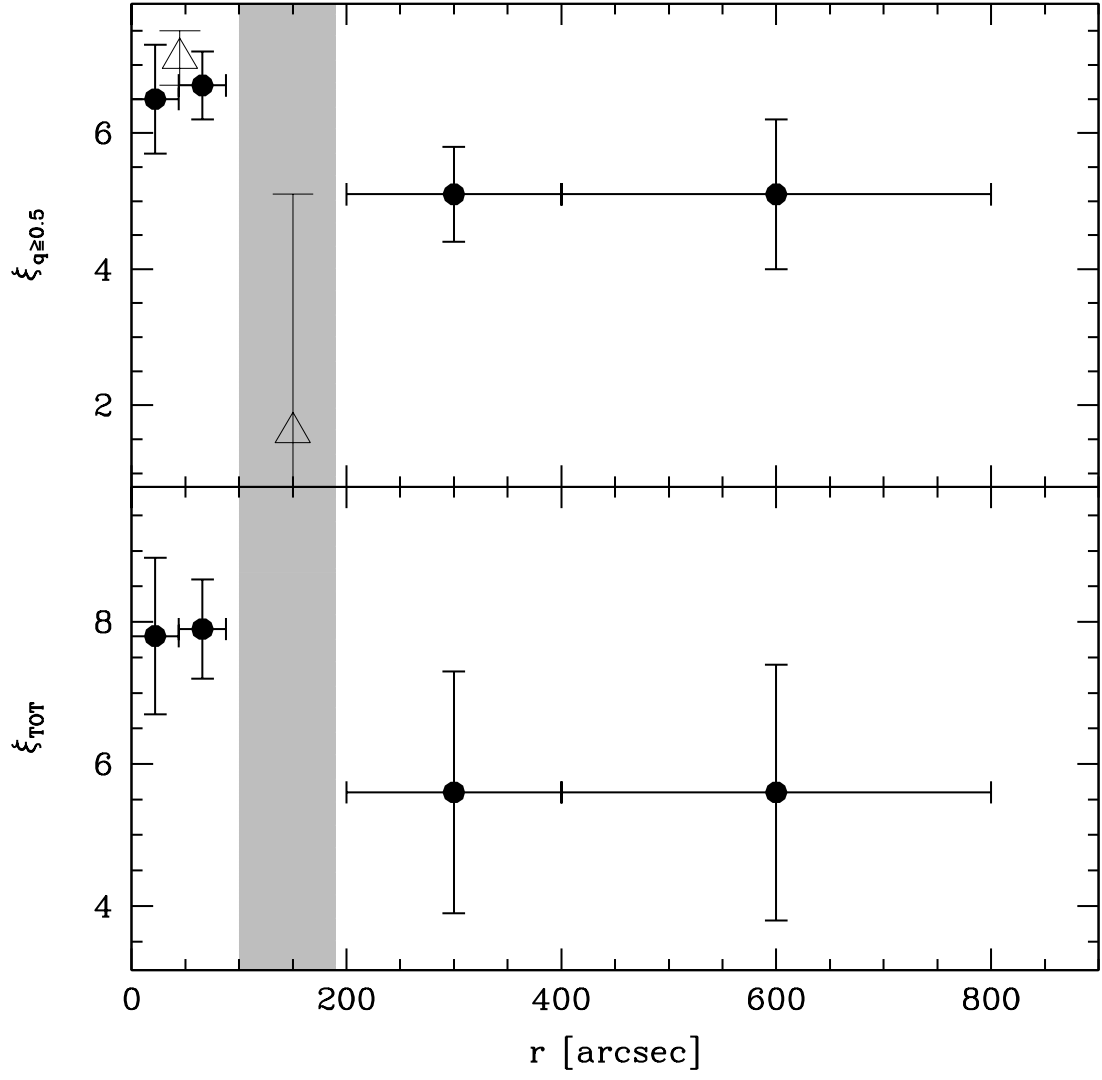}
\caption{Minimum and total binary fractions (solid
    circles in the upper and lower panels, respectively) as a function
    of radial distance from the cluster center. The grey area show the
    region excluded from the analysis because of the very low
    photometric completeness. The empty triangles in the upper panel
    mark the estimate from \citet{mil12}.}
\label{bin}
\end{figure}

\begin{figure}
\plotone{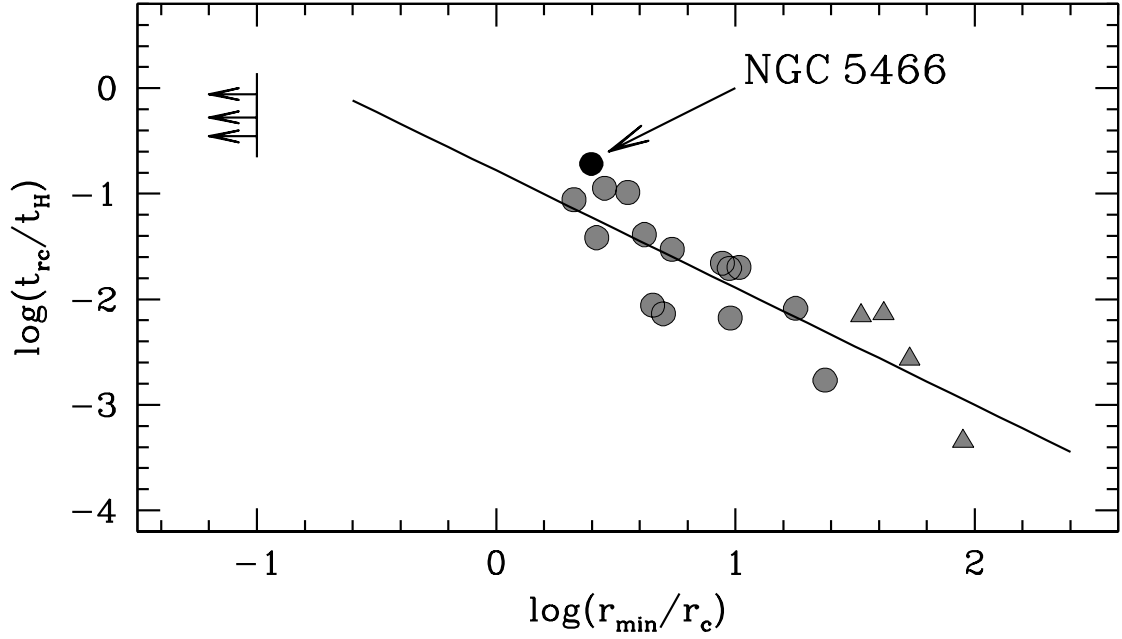} 
\caption{Core relaxation time ($t_{\rm rc}$) normalized to the age of
  the Universe ($t_{\rm H}$ = 13.7 Gyr), as a function of $r_{\rm
    min}/r_c$ (with $r_{\rm min}=180\arcsec$ being the position of the
  minimum of the BSS radial distribution). The figure is the same as
  Figure 4 in F12: the arrows indicate the location of the dynamically
  young clusters (\emph{Family I}); grey circles are GCs of
  intermediate age (\emph{Family II} in F12, plus M10 studied by
  \citealp{dale13}); grey triangles are dynamically old clusters
  (\emph{Family III}). The location of NGC 5466 in this plane is
  marked with the black circle and clearly shows that this is the
  second youngest member of \emph{Family III}.}
\label{clock}
\end{figure}

The central values of the derived binary fractions ($\xi_{q\ge
  0.5}=6.5\%$ and $\xi_{\rm TOT}=8\%$) are slightly smaller than those
quoted by \citet{sol07}. This is due to the differences in the adopted
data reduction procedures and the consequent completeness analysis
results, as well as the assumed luminosity functions.  The value of
$\xi_{q\ge 0.5}$ is in very good agreement with the estimate obtained
in the same radial range by \citet{mil12} from the same ACS
  data-sets: $\xi_{q\ge 0.5}=(7.1\pm 0.4)\%$ (see their Figure 34 and
  the innermost open triangle in Fig. \ref{bin}). Instead, our total
binary fraction is significantly lower than that quoted by
\citet[][$\xi_{\rm TOT}=14.2\%$]{mil12} because of the different
assumptions made about the shape of $f(q)$.  In fact, these authors assumed
a constant mass-ratio distribution for all $q$ values, and their total
binary fraction is simply twice the value of $\xi_{q\ge 0.5}$. Indeed,
the measure of $\xi_{\rm TOT}$ is quite sensitive to the assumed
mass-ratio distribution, as also demonstrated by the value (11.7\%)
obtained by simulating binaries formed by random associations between
stars of different masses \citep{sol07}.

 \citet{mil12} also provides the value of the minmum binary
  fraction abetween $2.35'$ and $2.45'$ from the cluster center:
  $\xi_{q\ge 0.5} = (1.6\pm 3.5)$\% . We emphasize, however, that this
  estimate has been obtained by using only the most external fragments
  of the ACS FoV, corresponding to less than 5\% of the total sampled
  area (this likely explains the large uncertainty quoted by the
  authors).  Keeping this caveat in mind, in light of the good
  agreement between our central value of $\xi_{q\ge 0.5}$ and that
  measured by \citet{mil12}, we include their estimate in our
  analysis, thus sampling the intermediate region which is not covered
  by our data.  Very interestingly, the value computed by
  \citet{mil12} defines a minimum in the binary fraction radial
  distribution (see the outer open triangle in Fig. \ref{bin}), thus
  implying that also the radial distribution of binaries seems to be
  bimodal in NGC 5466. We stress that the investigated binary systems
  (having primary star masses between 0.5 and $0.74 M_\odot$) are in a
  mass range consistent with that of BSSs. Indeed, within the
  uncertainties, also the position of the minimum of the binary radial
  distribution is in good agreement with that of the BSS population.
  This result urges a confirmation through dedicated observations.  In
  fact, we emphasize that this would be the first time that a similar
  feature is observed in a GC. It would further strengthen the
  interpretation proposed by F12 that the shape of the BSS radial
  distribution (and simialr mass objects) is primarily sculpted by
  dynamical friction.  Moreover, it adds further support to the
  conclusions that the unperturbed evolution of primordial binaries
  could be the dominant BSS formation process in low-density
  environments \citep{sol08}.

\acknowledgments{This research is part of the project COSMIC-LAB
  funded by the European Research Council (under contract
  ERC-2010-AdG-267675). GB acknowledges the European Community's
  Seventh Framework Programme under grant agreement no. 229517.  Based
  on observations obtained with MegaPrime/MegaCam, a joint project of
  CFHT and CEA/DAPNIA, at the Canada-France-Hawaii Telescope (CFHT)
  which is operated by the National Research Council (NRC) of Canada,
  the Institut National des Science de l'Univers of the Centre
  National de la Recherche Scientifique (CNRS) of France, and the
  University of Hawaii. Also based on observations made with the
  NASA/ESA Hubble Space Telescope, obtained from the data archive at
  the Space Telescope Institute. STScI is operated by the association
  of Universities for Research in Astronomy, Inc. under the NASA
  contract NAS 5-26555.  }

\clearpage

%% Use the figure environment and \plotone or \plottwo to include
%% figures and captions in your electronic submission.
%% To embed the sample graphics in
%% the file, uncomment the \plotone, \plottwo, and
%% \includegraphics commands
%%
%% If you need a layout that cannot be achieved with \plotone or
%% \plottwo, you can invoke the graphicx package directly with the
%% \includegraphics command or use \plotfiddle. For more information,
%% please see the tutorial on "Using Electronic Art with AASTeX" in the
%% documentation section at the AASTeX Web site,
%% http://www.journals.uchicago.edu/AAS/AASTeX.
%%
%% The examples below also include sample markup for submission of
%% supplemental electronic materials. As always, be sure to check
%% the instructions to authors for the journal you are submitting to
%% for specific submissions guidelines as they vary from
%% journal to journal.

%% This example uses \plotone to include an EPS file scaled to
%% 80% of its natural size with \epsscale. Its caption
%% has been written to indicate that additional figure parts will be
%% available in the electronic journal.

%\begin{figure}
%\plotone{map.ps} 
%\caption{\label{fig_map} \footnotesize{Map of the combined data set used to sample the total radial extension of NGC5466. 
%The dashed circle defines the cluster area probed by the LBC-Blue data, while the outer square corresponds to 
%the MEGACAM field of view. The solid circle corresponds to the location of the cluster's tidal radius.}}
%\end{figure} 

\clearpage

%% Here we use \plottwo to present two versions of the same figure,
%% one in black and white for print the other in RGB color
%% for online presentation. Note that the caption indicates
%% that a color version of the figure will be available online.
%%

%% If you are not including electonic art with your submission, you may
%% mark up your captions using the \figcaption command. See the
%% User Guide for details.
%%
%% No more than seven \figcaption commands are allowed per page,
%% so if you have more than seven captions, insert a \clearpage
%% after every seventh one.

%% Tables should be submitted one per page, so put a \clearpage before
%% each one.

%% Two options are available to the author for producing tables:  the
%% deluxetable environment provided by the AASTeX package or the LaTeX
%% table environment.  Use of deluxetable is preferred.
%%

%% Three table samples follow, two marked up in the deluxetable environment,
%% one marked up as a LaTeX table.

%% In this first example, note that the \tabletypesize{}
%% command has been used to reduce the font size of the table.
%% We also use the \rotate command to rotate the table to
%% landscape orientation since it is very wide even at the
%% reduced font size.
%%
%% Note also that the \label command needs to be placed
%% inside the \tablecaption.

%% This table also includes a table comment indicating that the full
%% version will be available in machine-readable format in the electronic
%% edition.

\clearpage
\begin{table}
\label{tab_obs}
 \centering
  \caption{Log of the observations}
  \begin{tabular}{@{}ccccc@{}}
  \hline
   Data-set     & Number of exposures & Filter & Exposure time & Date of observations\\
            &                 &        &     (s) & \\
 \hline
\multicolumn{5}{c}{Shallow Sample}\\
%  &   & Shallow Sample   &  \\           
 \hline
  LBC-blue   & 1 & $B$ & 5& 2007-06-17\\
      & 7 & $B$  & 90 & 2007-06-17\\
     & 1 & $V$ & 5 & 2007-06-17\\
      & 7 & $V$  & 60 & 2007-06-17\\
 MEGACAM & 1 &  $g$ & 90 & 2004-04-14\\
  & 1 & $r$ & 180 & 2004-04-14\\
 \hline
\multicolumn{5}{c}{Deep Sample}\\
 \hline
 LBC-blue  & 11 & $B$ & 400 & 2010-04-11\\
      & 15 & $V$ & 200 & 2010-04-11\\
  ACS   & 5 & $V_{606}$ & 340 & 2006-04-12\\
       & 5 & $I_{814}$ & 350 & 2006-04-12\\
\hline
\end{tabular}
\end{table}

%%%%%%%%%%%
\begin{table}
\label{tab_bss}
\centering
\caption{BSS radial distribution in NGC 5466}
  \begin{tabular}{@{}ccccc@{}}
  \hline
   $r_i\arcsec$     & $r_e\arcsec$ & $N_{\rm BSS}$ & $N_{\rm RGB}$ & Fraction of cluster's luminosity\\
 \hline
           0      &   72    &     32 (0.01)   &    295 (0.05)  &      0.211       \\
         72    &     150   &      37 (0.02)  &     510 (0.17) &      0.353     \\
         150    &     200   &       4  (0.02)  &     204 (0.17) &     0.136         \\
         200    &     500   &       11 (0.27)    &     321 (2.01) &    0.230      \\
         500    &    1400   &       10 (2.16)  &      109  (16.36)  &        0.07   \\ 	
 \hline
\end{tabular}
\tablecomments{Number of BSSs and RGB stars, and  fraction of sampled light in the radial annuli considered to study the BSS radial distribution. The field contamination expected for each population is reported in parentheses.}
\end{table}

%%%%%%%%%%%
\begin{table}
\label{tab_bina}
\centering
\caption{Binary fraction of NGC 5466}
\begin{tabular}{@{}cccccc@{}}
\hline
   $r_i\arcsec$     & $r_e\arcsec$ & $\xi_{min}$ &  $\xi_{TOT}$  \\
\hline
           0      &   44    &     ($6.5\pm0.8$)\%  &     ($7.8\pm1.1$)\%       \\
         44    &     88   &     ($6.7\pm0.5$)\%  &     ($7.9\pm0.7 $)\%    \\
         200    &     400   &    ($5.1\pm0.7$)\% &  ($5.6\pm1.7 $)\%    \\
         400    &   800   &       ($5.1\pm1.1$)\%  &    ($ 5.6\pm1.8$)\%  \\ 	
\hline
\end{tabular}
\tablecomments{Minimum ($\xi_{min}$) and total ($\xi_{TOT}$) binary fraction as calculated in 4 concentric annuli.}
\end{table}


\begin{thebibliography}{}

\bibitem[Bailyn(1992)]{ba92} Bailyn, C.~D.\ 1992, \apj, 392, 519

\bibitem[Bailyn(1995)]{ba95} Bailyn, C.~D.\ 1995, \araa, 33, 133

\bibitem[Beccari et al.(2008)]{be08} Beccari, G., Lanzoni, 
B., Ferraro, F.~R., et al.\ 2008, \apj, 679, 712 

\bibitem[Beccari et al.(2011)]{be11} Beccari, G., Sollima, 
A., Ferraro, F.~R., et al.\ 2011, \apjl, 737, L3 

\bibitem[Beccari et al.(2012)]{be12} Beccari, G., 
L{\"u}tzgendorf, N., Olczak, C., et al.\ 2012, \apj, 754, 108 

%\bibitem[Bedin et al.(2005)]{bed05} Bedin, L.~R., Cassisi, S., Castelli, F.,
%Piotto, G., Anderson, J., Salaris, M., Momany, Y., \& Pietrinferni, A.\ 2005,
%\mnras, 357, 1038

\bibitem[Bellazzini et al.(1995)]{bellazz95} Bellazzini, M., Pasquali,
  A., Federici, L., Ferraro, F.~R., \& Pecci, F.~F.\ 1995, \apj, 439,
  687

\bibitem[Bellazzini et al.(2002)]{be02} Bellazzini, M., Fusi Pecci, F.,
Messineo, M., Monaco, L., \& Rood, R.~T.\ 2002, \aj, 123, 1509

\bibitem[Belokurov et al.(2006)]{bel06} Belokurov, V., Evans, 
N.~W., Irwin, M.~J., Hewett, P.~C., \& Wilkinson, M.~I.\ 2006, \apjl, 637, L29 

\bibitem[Bertin \& Arnouts(1996)]{BA96} Bertin, E., \& Arnouts, S.\ 1996, \aaps, 117, 393 

%\bibitem[Buonanno et al.(1984)]{buon84} Buonanno, R., Buscema, 
%G., Corsi, C.~E., Iannicola, G., \& Fusi Pecci, F.\ 1984, \aaps, 56, 79 

%\bibitem[Clement et al.(2001)]{clem01} Clement, C.~M., et al.\ 2001, \aj, 122, 2587

\bibitem[Contreras Ramos et al.(2012)]{con12} Contreras 
Ramos, R., Ferraro, F.~R., Dalessandro, E., Lanzoni, B., 
\& Rood, R.~T.\ 2012, \apj, 748, 91 

\bibitem[Cote et al.(1996)]{co96} Cote, P., Pryor, C., McClure, R.~D., Fletcher, J.~M., \& Hesser, J.~E.\ 1996, \aj, 112, 574 

\bibitem[Dalessandro et al.(2008)]{da08} Dalessandro, E., 
Lanzoni, B., Ferraro, F.~R., et al.\ 2008, \apj, 681, 311 

\bibitem[Dalessandro et al.(2009)]{da09} Dalessandro, E., 
Beccari, G., Lanzoni, B., et al.\ 2009, \apjs, 182, 509 

\bibitem[Dalessandro et al.(2011)]{da11} Dalessandro, E., 
Lanzoni, B., Beccari, G., et al.\ 2011, \apj, 743, 11 

\bibitem[Dalessandro et al.(2013)]{dale13} Dalessandro, E., Ferraro,
  F.~R., Lanzoni, B., et al.\ 2013, \apj, 770, 45

\bibitem[Djorgovski(1993)]{dj93} Djorgovski, S. 1993, in ASPC Conf. Ser. 50, Structure and Dynamics of Globular Clusters, ed. S. G. Djorgovski \& G. Meylan (San Francisco: ASP), 373D

\bibitem[Dotter et al.(2007)]{dot07} Dotter, A., Chaboyer, 
B., Jevremovi{\'c}, D., et al.\ 2007, \aj, 134, 376 

\bibitem[Fekadu et al.(2007)]{fek07} Fekadu, N., Sandquist, 
E.~L., \& Bolte, M.\ 2007, \apj, 663, 277 

\bibitem[Ferraro et al.(1991)]{fe91} Ferraro, F.~R., Clementini, G.,
  Fusi Pecci, F., \& Buonanno, R.\ 1991, \mnras, 252, 357

\bibitem[Ferraro et al.(1992)]{fe92} Ferraro, F.~R., 
  Fusi Pecci, F., \& Buonanno, R.\ 1992, \mnras, 256, 376

\bibitem[Ferraro et al.(1995)]{fe95} Ferraro, F. R., Fusi Pecci, F.,
  \& Bellazzini, M.\ 1995, \aap, 294, 80

\bibitem[Ferraro et al.(1997)]{f97} Ferraro, F.~R., Paltrinieri, B., Fusi Pecci, F., et al.\ 1997, \aap, 324, 915 

\bibitem[Ferraro et al.(1999)]{f99} Ferraro, F.~R., 
Messineo, M., Fusi Pecci, F., et al.\ 1999, \aj, 118, 1738 

\bibitem[Ferraro et al.(2003)]{fe03} Ferraro, F.~R., Sills, 
A., Rood, R.~T., Paltrinieri, B., \& Buonanno, R.\ 2003, \apj, 588, 464 

\bibitem[Ferraro et al.(2004)]{f04} Ferraro, F.~R., 
Beccari, G., Rood, R.~T., et al.\ 2004, \apj, 603, 127 


\bibitem[Ferraro et al.(2006a)]{fe06_COdep} Ferraro, F.~R., Sabbi, 
E., Gratton, R., et al.\ 2006a, \apjl, 647, L53 

\bibitem[Ferraro et al.(2006b)]{fe06_ocen} Ferraro, F.~R., 
Sollima, A., Rood, R.~T., et al.\ 2006b, \apj, 638, 433 

\bibitem[Ferraro et al.(2009)]{fe09_m30} Ferraro, F.~R., 
Beccari, G., Dalessandro, E., et al.\ 2009, \nat, 462, 1028 (F09)

\bibitem[Ferraro et al.(2012)]{fe12} Ferraro, F.~R., Lanzoni, B., Dalessandro, E., et al.\ 2012, \nat, 492, 393 

\bibitem[Fisher et al.(2005)]{fi05} Fisher, J., Schr{\"o}der, K.-P., \& Smith, R.~C.\ 2005, \mnras, 361, 495 

\bibitem[Giallongo et al.(2008)]{gial08} Giallongo, E., Ragazzoni, R., Grazian, A., et al.\ 2008, \aap, 482, 349 

\bibitem[Gilliland et al.(1998)]{gilliland98} Gilliland, R.~L., Bono,
  G., Edmonds, P.~D., et al.\ 1998, \apj, 507, 818

\bibitem[Goldsbury et al.(2010)]{g10} Goldsbury, R., 
Richer, H.~B., Anderson, J., et al.\ 2010, \aj, 140, 1830 

\bibitem[Grillmair \& Johnson(2006)]{gr06} Grillmair, C.~J., \& Johnson, R.\ 2006, \apjl, 639, L17 

\bibitem[Harris(1996)]{ha96} Harris, W.~E.\ 1996, \aj, 112, 
1487

\bibitem[Hill et al.(2006)]{hill06} Hill, J.~M., Green, R.~F., \& Slagle, J.~H.\ 2006, \procspie, 6267, 62670Y

\bibitem[Hills \& Day (1976)]{hillsday76} Hills, J. G., \& Day,
  C. A.\ 1976, Astrophys. Lett., 17, 87

\bibitem[Hurley \& Tout(1998)]{hu98} Hurley, J., \& Tout, C.~A.\ 1998, \mnras, 300, 977 

\bibitem[Hut et al.(1992)]{hut92} Hut, P., McMillan, S., Goodman, J.,
  et al.\ 1992, \pasp, 104, 981

\bibitem[King(1966)]{k66} King, I.R. 1966, AJ, 71, 64

\bibitem[Knigge et al.(2009)]{kn09} Knigge, C., Leigh, N., \& Sills, A.\ 2009, \nat, 457, 288 

\bibitem[Kroupa(2002)]{kr02} Kroupa, P.\ 2002, Science, 295, 82 

\bibitem[Kryachko et al.(2011)]{kr11} Kryachko, T., 
Samokhvalov, A., \& Satovskiy, B.\ 2011, Peremennye Zvezdy Prilozhenie, 11, 20 

\bibitem[Lanzoni et al.(2007a)]{lanz07_1904} Lanzoni, B., Sanna, N., 
Ferraro, F.~R., et al.\ 2007a, \apj, 663, 1040

\bibitem[Lanzoni et al.(2007b)]{lanz07_M5} Lanzoni, B., 
Dalessandro, E., Ferraro, F.~R., et al.\ 2007b, \apj, 663, 267 

%\bibitem[Lehmann \& Scholz(1997)]{leh97} Lehmann, I., \& Scholz, R.-D.\ 1997, \aap, 320, 776 

%\bibitem[Leigh et al.(2011)]{le11} Leigh, N., Sills, A., \& Knigge, C.\ 2011b, \mnras, 415, 3771

\bibitem[Leigh et al.(2013)]{leigh13} Leigh, N., Knigge, C., 
Sills, A., et al.\ 2013, \mnras, 428, 897 

%\bibitem[Mapelli et al.(2004)]{map04} Mapelli M., Sigurdsson S., Colpi M., Ferraro F. R., Possenti A., Rood R. T., Sills A., Beccari G., 2004, ApJ, 605, L29
%
%\bibitem[Mapelli et al.(2006)]{map06} Mapelli M., Sigurdsson S., Ferraro F. R., Colpi M., Possenti A., Lanzoni B., 2006, MNRAS, 373, 361

\bibitem[Mathieu \& Geller(2009)]{math09} Mathieu, R.~D., \& Geller, A.~M.\ 2009, \nat, 462, 1032 

\bibitem[Mateo et al.(1990)]{ma90} Mateo, M., Harris, H.~C., 
Nemec, J., \& Olszewski, E.~W.\ 1990, \aj, 100, 469 

\bibitem[Mateo(1996)]{mat96} Mateo, M.\ 1996, The Origins, Evolution, and Destinies of Binary Stars in Clusters, ASPCS, 90, 21 

\bibitem[McCrea (1964)]{mcrea64} McCrea, W. H.\ 1964, \mnras, 128, 147

\bibitem[McMillan(1991)]{mc91} McMillan, S.~L.~W.\ 1991, The 
Formation and Evolution of Star Clusters, 13, 324 

\bibitem[Meylan \& Heggie(1997)]{me97} Meylan, G., \& 
Heggie, D.~C.\ 1997, \aapr, 8, 1 

\bibitem[Milone et al.(2012)]{mil12} Milone, A.~P., Piotto, G., Bedin, L.~R., et al.\ 2012, \aap, 540, A16 

\bibitem[Miocchi et al.(2013)]{mio13} Miocchi, P., Lanzoni, 
B., Ferraro, F.~R., et al.\ 2013, arXiv:1307.6035 

%\bibitem[Moretti et al.(2008)]{mo08} Moretti, A., de Angeli, F., \& Piotto, G.\ 2008, \aap, 483, 183 

\bibitem[Nemec \& Harris(1987)]{ne87} Nemec, J.~M., \& 
Harris, H.~C.\ 1987, \apj, 316, 172 

\bibitem[Paresce et al.(1992)]{paresce92} Paresce, F., de Marchi, G.,
  \& Ferraro, F.~R.\ 1992, \nat, 360, 46

\bibitem[Pooley \& Hut(2006)]{pooley06} Pooley, D., \& Hut, P.\ 2006,
  \apjl, 646, L143

\bibitem[Ragazzoni et al.(2006)]{rag06} Ragazzoni, R., et al.\ 2006, \procspie, 6267, 626710

\bibitem[Ransom et al.(2005)]{ransom05} Ransom, S.~M., Hessels,
  J.~W.~T., Stairs, I.~H., et al.\ 2005, Science, 307, 892

\bibitem[Renzini \& Buzzoni(1986)]{ren89} Renzini A., Buzzoni A., 1986, ASSL, 122, 195

\bibitem[Rey et al.(1998)]{r98} Rey, S.-C., Lee, Y.-W., 
Byun, Y.-I., \& Chun, M.-S.\ 1998, \aj, 116, 1775 

\bibitem[Robin et al.(2003)]{rob03} Robin, A.~C., Reyl{\'e}, 
C., Derri{\`e}re, S., \& Picaud, S.\ 2003, \aap, 409, 523 

\bibitem[Romani \& Weinberg(1991)]{ro91} Romani, R.~W., \& Weinberg, M.~D.\ 1991, \apj, 372, 487 

\bibitem[Sarajedini et al.(2007)]{sa07} Sarajedini, A., 
Bedin, L.~R., Chaboyer, B., et al.\ 2007, \aj, 133, 1658 

%\bibitem[Schechter, Mateo \& Saha (1993)]{dophot} Schechter, P. L., Mateo, M.,
%\& Saha, A.\ 1993, PASP, 105, 1342

\bibitem[Shara et al. (1997)]{shara97} Shara, M. M., Saffer, R. A., \&
  Livio, M.\ 1997, \apj, 489, L59

\bibitem[Sirianni et al.(2005)]{sir05} Sirianni, M., et al.\ 2005,
PASP, 117, 1049 

\bibitem[Sollima et al.(2007)]{sol07} Sollima, A., Beccari, 
G., Ferraro, F.~R., Fusi Pecci, F., \& Sarajedini, A.\ 2007, \mnras, 380, 
781 
\bibitem[Sollima et al. (2008)]{sol08} Sollima, A. et al.\ 2008,
  \aap, 481, 701

\bibitem[Sollima et al.(2010)]{sol10} Sollima, A., Carballo-Bello, J.~A., Beccari, G., et al.\ 2010, \mnras, 401, 577 

\bibitem[Stetson(1987)]{st87}Stetson, P. B. 1987, \pasp, 99, 191

\bibitem[Stetson(1994)]{st94}Stetson, P. B. 1994, \pasp, 106, 250

\bibitem[Stetson(2000)]{ste00} Stetson, P.~B.\ 2000, \pasp, 112, 925 

\bibitem[Wilson(1975)]{w75}Wilson, C. P. 1975, AJ, 80, 175

\bibitem[Zinn \& Searle (1976)]{zinnsearle76} Zinn, R., \& Searle, L.\ 1976,
  \apj, 209, 734

\end{thebibliography}
\end{document}